\documentclass[]{piparticle-final}
\usepackage{graphicx}
\usepackage{amsmath}
\usepackage{subfigure}
\usepackage{psfrag}

\usepackage{cite} % To compress citation ranges
\usepackage{epstopdf} 

\begin{document}

\volume{6}               % To be inserted by Editor
\articlenumber{060013}   % To be inserted by Editor
\journalyear{2014}       % To be inserted by Editor
\editor{G. C. Barker}   % To be inserted by Editor
%\reviewers{Reviewer's name}  % To be inserted by Editor
\received{9 October 2014}     % To be inserted by Editor
\accepted{13 November 2014}   % To be inserted by Editor
\runningauthor{D. R. Parisi \itshape{et al.}}  % To be inserted by Editor
\doi{060013}         % To be inserted by Editor

\title{Sequential evacuation strategy for multiple rooms toward the same means of egress}

% Institution references with \cite are inserted after \maketitle in theaffiliation enviroment
\author{D. R. Parisi,\cite{inst1,inst2}\thanks{E-mail: dparisi@itba.edu.ar}  \hspace{0.5em}
        P. A. Negri\cite{inst2,inst3}\thanks{E-mail: pnegri@uade.edu.ar}}

\pipabstract{
This paper examines different evacuation strategies for systems where several rooms evacuate  through the same means of egress, using microscopic pedestrian simulation.
As a case study, a medium-rise office building is considered. It was found that the standard strategy, whereby the simultaneous evacuation of all levels is performed, can be improved by a sequential evacuation, beginning with the  lowest floor and continuing successively with each one of the upper floors after a certain delay. The importance of the present research is that it provides the basis for the design and implementation of new evacuation strategies and alarm systems that could significantly improve the evacuation of multiple rooms  through a common means of escape. 
%\\
%\textbf{Keyworks}: pedestrian dynamics, evacuation simulation, egress strategy, multistory buildings. 
}

\maketitle

\blfootnote{
\begin{theaffiliation}{99}
   \institution{inst1} Instituto Tecnol\'{o}gico de Buenos Aires,  25 de Mayo 444, 1002 Ciudad Aut\'{o}noma de Buenos Aires, Argentina.
   \institution{inst2} Consejo Nacional de Investigaciones Cient\'{i}ficas y T\'{e}cnicas, Av. Rivadavia 1917, 1033 Ciudad Aut\'{o}noma de Buenos Aires, Argentina.
   \institution{inst3} Universidad Argentina de la Empresa, Lima 754, 1073 Ciudad Aut\'{o}noma de Buenos Aires, Argentina.
\end{theaffiliation}
}

%%%%%%%%%%%%%%%%%%%%%%%%%%%%%%%%%%%%%%%%%%%%%%%%%%%%%%%%%%%%%%%%%%%%%%%%%%%%%%%%%%%%%%%%%%%%5
\section{Introduction}\label{sec:introduction}

A quick and safe evacuation of a building when threats or hazards are present, whether natural or man-made, is of enormous interest in the field of safety design.
Any improvement in this sense would increase evacuation safety, and a greater number of lives could be better protected when fast and efficient total egress  is required. 

Evacuation from real pedestrian facilities can have different degrees of complexity due to the particular layout, functionality, means of escape, occupation and evacuation plans. During the last two decades, modeling and simulation of pedestrian movements have developed into a new approach to the study of this kind of system. Basic research on evacuation dynamics  has started with the simplest problem  of evacuation from a room through a single door. This ``building block'' problem of pedestrian evacuation has extensively been studied in the bibliography, for example, experimetally \cite{Kretz:2006, Seyfried:2009}, or by using the social force model \cite{Helbing:2000, Parisi:2005, Parisi:2007}, and cellular automata models \cite{Kirchner:2002,Burstedde:2001,Song:2006}, among many others.

As a next step, we propose  investigating the egress from multiple rooms toward a single means of egress,  such as a hallway or corridor. Examples of this configuration are schools and universities where several classrooms  open into a single hallway, cinema complexes, museums, office buildings, and the evacuation of different building floors via the same staircase. The key variable in this kind of system is the timing (simultaneity) at which the different occupants of individual rooms go toward the common means of egress. Clearly, this means of egress has a certain capacity  that can be rapidly exceeded if all rooms are evacuated simultaneously and thus, the total evacuation time can be suboptimal. So, it is valid to ask in what order the different rooms should be evacuated.

The answer to this question is not obvious. Depending on the synchronization and order in which the individual rooms are evacuated, the hallway can be saturated in different sectors, which could hinder the exit  from some rooms and thus, the corresponding flow rate of people will be limited by the degree of saturation of the hallway. This is because  density is a limitation for speed. The relationship between density and velocity in a crowd is called ``fundamental diagram of pedestrian traffic'' \cite{Weidmann:1993,Fruin:1971,Seyfried:2005,DiNenno:2002,Helbing:2007,pednet}. Therefore, the performance of the egress from each room will depend on the density of people  in the hallway,  which is difficult to predict from analytical methods. This type of analysis is limited to simple cases such as simultaneous evacuation of all rooms, assuming a maximum degree of saturation on the stairs. An example of an analytical resolution for this simple case can be seen in Ref. 
\cite{DiNenno:2002}, on chapter 3-14, where the 
egress from a multistory building is studied. 

From now on we will analyze a 2D version of this particular case: an office building with 7 floors being evacuated through the same staircase, which is just an example of the general problem of several rooms evacuating through a common means of egress. 

\subsection{Description of the evacuation process}\label{subsec:DescEvac}

The evacuation process comprises two periods: 
\begin{itemize}
\item[-] $E_1$, reaction time indicating the time period between the onset of a threat or incident and the instant when the occupants of the building begin to evacuate.
\item[-] $E_2$, the evacuation time itself is measured from the beginning of the egress, when the first person starts to exit, until the last person is able leave the building. 
\end{itemize}

$E_1$ can be subdivided into: time to detect danger, report to building manager, decision-making of the person responsible for starting the evacuation, and the time it takes to activate the alarm. 
These times are of variable duration depending on the usage given to the building, the day and time of the event, the occupants training, the proper functioning of the alarm system, etc. 
Because period $E_1$ takes place before the alarm system is triggered, it must be separated from period $E_2$. The duration of $E_1$ is the same for the whole building. In consequence, for the present study only the evacuation process itself described as period $E_2$ is considered. 
The total time of a real complete evacuation will be necessarily longer depending on the duration of $E_1$. 

\subsection{Hypothesis}

This subsection defines the scope and conditions that are assumed for the system. 
\begin{enumerate}
\item The study only considers period $E_2$ (the evacuation process itself) described in subsection \ref{sec:introduction} \ref{subsec:DescEvac} above.
\item All floors have the same priority for evacuation. The case in which there is a fire at some intermediate floor is not considered.
\item The main aspect to be analyzed is the movement of people who follow the evacuation plan. Other aspects of safety such as types of doors, materials, electrical installation, ventilation system, storage of toxic products, etc., are not included in the present analysis.
\item After the alarm is triggered on each floor, the egress begins under conditions similar to those of a fire drill, namely:
	\begin{itemize}
		\item People walk under normal conditions, without running.
		\item If high densities are produced, people wait without pushing.
		\item Exits are free and the doors are wide open. 
		\item The evacuation plan is properly signaled. 
		\item People start to evacuate when the alarm is activated on their own floor, following the evacuation signals.
		\item There is good visibility. 
	\end{itemize}
\end{enumerate}

%%%%%%%%%%%%%%%%%%%%%%%%%%%%%%%%%%%%%%%%%%%%%%%%%%%%%%%%%%%%%
\section{Simulations}\label{sec:simul}

\subsection{The model}

The physical model implemented is the one described in \cite{Parisi:2009}, which is a modification of the social force model (SFM) \cite{Helbing:2000}. This modification allows a better approximation to the fundamental diagram of Ref. \cite{DiNenno:2002}, commonly used in the design of pedestrian facilities.

The SFM is a continuous-space and force-based model that describes the dynamics considering the forces exerted over each particle ($p_{i}$). Its Newton equation reads

\begin{equation}
m_i \mathbf{a}_{i}= \mathbf{F}_{Di}+\mathbf{F}_{Si}+\mathbf{F}_{Ci},
\label{Newton}
\end{equation}
where $\mathbf{a}_{i}$ is the acceleration of particle $p_{i}$. The equations are solved using standard molecular dynamics techniques. The three
forces are: ``Driving Force'' ($\mathbf{F}_{Di} $), ``Social Force'' ($\mathbf{F}_{Si}$) and ``Contact Force''($\mathbf{F}_{Ci}$). The corresponding expressions are as follows

\begin{equation}
\mathbf{F}_{Di}=m_{i}~\frac{(v_{di}~\mathbf{e}_{i}-\mathbf{v}_{i})}{\tau },
\label{F_D}
\end{equation}
where $m_{i}$ is the particle mass, $\mathbf{v}_{i}$ and $v_{di}$ are the actual velocity and the desired velocity magnitude, respectively. $\mathbf{e}_{i}$ is the unit vector pointing to the desired target (particles inside the corridors or rooms have their targets located at the closest position over the line of the exit door), $\tau $ is a constant related to the time needed for the particle to achieve $v_{d}$.

\begin{equation}
\mathbf{F}_{Si}=\sum_{j=1,j\neq i}^{N_\text{p}}~A~\exp \left(\frac{-\epsilon _{ij}}{B}
\right)~\mathbf{e}_{ij}^{n},
\label{F_S}
\end{equation}
with $N_\text{p}$ being the total number of pedestrians in the system, $A$ and $B$
are constants that determine the strength and range of the social
interaction, $\mathbf{e}_{ij}^{n}$ is the unit vector pointing from particle 
$p_{j}$ to $p_{i}$; this direction is the ``normal'' direction between two
particles, and $\epsilon_{ij}$ is defined as

\begin{equation}
\epsilon _{ij}=r_{ij}-(R_{i}+R_{j}),
\end{equation}
where $r_{ij}$ is the distance between the centers of $p_{i}$ and $p_{j}$
and $R$ is their corresponding particle radius.

\begin{align}
&\mathbf{F}_{Ci}=\\
&\sum_{j=1,j\neq i}^{N_\text{p}}\left[ (-\epsilon_{ij}~k_{n})~\mathbf{e}_{ij}^{n}+({v}_{ij}^{t}~\epsilon_{ij}~k_{t})~\mathbf{e}_{ij}^{t}\right] ~g(\epsilon_{ij}),
\label{F_C}\notag
\end{align}
where the tangential unit vector ($\mathbf{e}_{ij}^{t}$) indicates the
corresponding perpendicular direction, $k_{n}$ and $k_{t}$ are the normal
and tangential elastic restorative constants, $v_{ij}^{t}$ is the tangential projection of the relative velocity seen from $p_{j}$($\mathbf{v}_{ij}=\mathbf{v}_{i}-\mathbf{v}_{j}$), and the function $g(\epsilon _{ij})$ is: $g=1$ if $\epsilon_{ij}<0 $ or $g=0$ otherwise. 

Because this version of the SFM does not provide any self-stopping mechanism for the particles, it  cannot reproduce the fundamental diagram of pedestrian traffic as shown in Ref. \cite{Parisi:2009}. In consequence, the modification consists in providing virtual pedestrians with a way to stop pushing other pedestrians. This is achieved by incorporating a semicircular respect area close to and ahead of the particle ($p_i$). While any other pedestrian is inside this area, the desired velocity of pedestrians ($p_i$) is set equal to zero ($v_{di}~=~0$). For further details and benefits of this modification to the SFM, we refer the reader to Ref. \cite{Parisi:2009}.

The kind of model used allows one to define the pedestrian characteristics individually. Following standard pedestrian dynamics bibliography (see, for example, \cite{Helbing:2000, Parisi:2005, Parisi:2007, Parisi:2009}), we considered independent and uniform distributed values between the ranges: pedestrian mass $m~\epsilon~$[70~kg,~90~kg]; shoulder width $d~\epsilon~$[48~cm,~56~cm]; desired velocity $~v_d~\epsilon~$[1.1~m/s,~1.5~m/s]; and the constant values are: $\tau=0.5$~s, $A=2000$~N, $B=0.08$~m, $k_n=1.2~10^5$~N/m, $k_t=2.4~10^5$~kg/m/s.

Beyond the microscopic model, pedestrian behavior simply consists in moving toward the exit of the room and then toward the exit of the hallway, following the evacuation plan.

From the simulations, all the positions and velocities of the virtual pedestrians were recorded every 0.1 second. From these data, it is possible to calculate several outputs; in the present work we focused on evacuation times.

\subsection{Definition of the system under study}\label{subsec:definitionSyste}

As a case study, we have chosen that of a medium-rise office building with $N=7$, $N$ being the number of floors. This system was studied analytically in Chapter 3-14 in Ref. \cite{DiNenno:2002}, only for the case of simultaneous evacuation of all floors.

The building has two fire escapes in a symmetric architecture. At each level, there are 300 occupants. Exploiting the symmetric configuration, we will only consider the egress of 150 persons toward one of the stairs. Thus, on each floor, 150 people are initially placed along the central corridor that is 1.2 m wide and 45 m long. In total, 1050 pedestrians are considered for simulating the system.

\begin{figure}%[th]
\begin{center}
\includegraphics[width=0.45\textwidth]{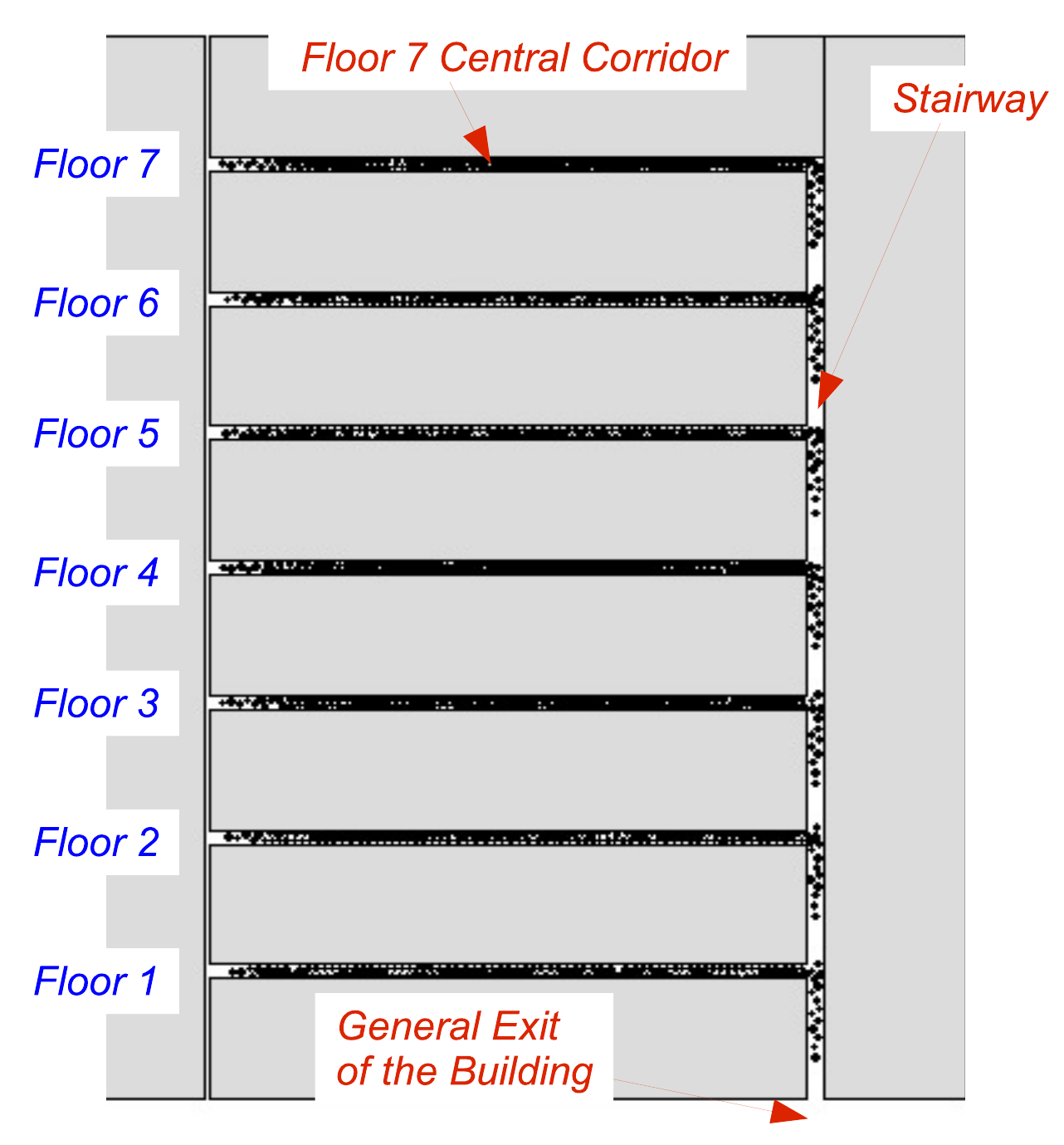}
\end{center}
\caption{Schematic of the two-dimensional system to be simulated. Each black dot indicates one person.} 
\label{figure2}
\end{figure}

For the sake of simplicity, we define a two-dimensional version of a building where the central corridors of all the floors and the staircase are considered to be  on the same plane as shown in Fig. \ref{figure2}. 

The central corridors can be identified with the ``rooms'' of the general problem described in section \ref{sec:introduction} and the staircase is the common means of egress. The effective width of the stairway is 1.4 m. 
The central corridors of each floor are separated by 10.66 m. This separation arises from adding the horizontal distance of the steps and the landings between floors in the 3D system \cite{DiNenno:2002}. So the distance between two floors in the 2D version of the problem is of the same length as the horizontal distance that a person should walk, also between two floors, along the stairway in the 3D building.

\subsection{Evacuation strategies}
The objective of proposing a strategy in which different floors start their evacuation at different times is to investigate whether this method allows an improvement over the standard procedure, which is the simultaneous evacuation of all floors. 

The parameters to be varied in the study are the following: 
\begin{itemize}
\item[a] The order in which the different levels are evacuated. In this sense, we study two procedures: a.1) ``Bottom-Up'':   indicates that the evacuation begins  on the lowest ($1^{st}$)  floor and then follows in order  to the immediately superior floors. a.2) ``Top-Down'' indicates that the evacuation begins  on the top floor ($7^{th}$, in this case), and continues to the next lower floor, until the $1^{st}$ floor is finally evacuated. 
\item[b] The time delay $dt$ between the start of the evacuation of two consecutive floors. This could be implemented in a real system through a segmented alarm system for each floor, which triggers the start of the evacuation in an independent way for the corresponding floor. 
\end{itemize}

The initial time, when the first fire alarm is triggered in the building, is defined as $T_0$.

The instant $t_{0~\{BU,TD,SE\}}^f$ indicates the time when the alarm is activated on floor $f$. Subindices $\{BU,TD,SE\}$ are set if the time $t$ belongs to the Bottom-Up, Top-Down, or Simultaneous Evacuation strategies, respectively.

The Bottom-Up strategy establishes that the $1^{st}$ floor is evacuated first: $t_{0~BU}^1=T_0$.
Then the alarm on the $2^{nd}$ floor  is triggered after $dt$ seconds, $t_{0~BU}^2 = t_{0~BU}^1+dt$, and so on in ascending order up to the $7^{th}$ floor .  
In general, the time when the alarm is triggered  on floor $f$ can be calculated as:

\begin{equation}
t_{0~BU}^f = T_0 + dt \times (f-1).\end{equation}

The Top-Down strategy begins the building evacuation  on the top floor ($7^{th}$, in this case): $t_{0~TD}^7=T_0$. After a time $dt$, the evacuation of the floor immediately below starts, and so on until the evacuation of the $1^{st}$ floor:

\begin{equation}
t_{0~TD}^f = T_0 + dt \times (N - f).\end{equation}

Simultaneous Evacuation is the special case in which $dt=0$ and thus, it considers the alarms  on all the floors to be triggered at the same time:  \begin{equation}
t_{0~SE}^f =T_0|_{f=1,2,...,7}.\end{equation}

%%%%%%%%%%%%%%%%%%%%%%%%%%%%%%%%%%%%%%%%%%%%%%%%%%%%%%%%%%%%%%%%%
\section{Results}\label{sec:result}
This section presents the results of simulations made by varying the strategy and the time delay between the beginning of the evacuation of the different levels.

Each configuration was simulated five times, and thus, the mean values and standard deviations are reported.
This is consistent with reality, because if a drill is repeated in the same building, total evacuation times will not be exactly the same.

\subsection{Metrics definition}\label{subsec:definitions}

Here we define the metrics that will be used to quantify the efficiency of the evacuation process of the system under study.

It is called \textit{Total Evacuation Time (TET)}, starting at $T_0$, when everyone in the building ($150 \times 7 = 1050$ persons)  has reached the exit located on the ground floor (see Fig. \ref{figure2}), which means that the building is completely evacuated.

The $f^{th}$ \textit{Floor Evacuation Time} ($FET_f$) refers to the time elapsed since initiating the evacuation of floor $f$ until its 150 occupants reach the staircase. 
It must be noted that this evacuation time does not consider the time elapsed between the access to the staircase  and the general exit  from the building, nor does it consider as starting time the time at which  the evacuation of some other level or of the building in general begins. It only considers the beginning of the evacuation of the current floor.
Average Floor Evacuation Time (${FET}$) is the average of the seven $FET_f$.

From these definitions, it  follows that $TET > FET_f$ for any floor (even the lowest one).

\subsection{Simultaneous evacuation strategy}\label{subsec:SimultaneousEvac}

\begin{figure}%[th]
\begin{center}
\includegraphics[width=0.45\textwidth]{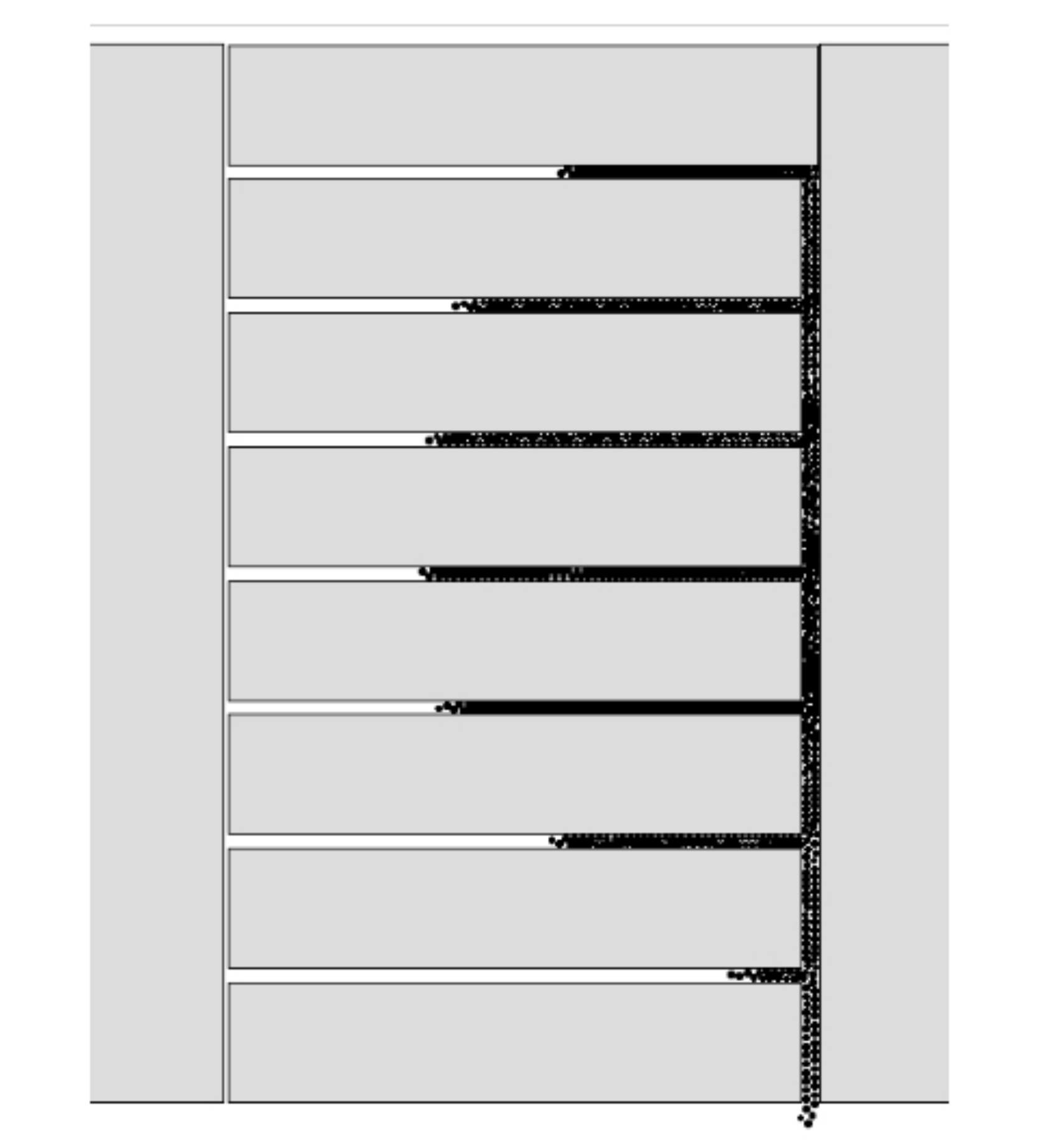}
\end{center}
\caption{Snapshot taken at 73 seconds since the start of the simultaneous evacuation, where the queues of different lengths can be observed on each floor.}  \label{figure3}
\end{figure}

In general, the standard methodology consists in evacuating all the floors having the same priority at the same time.

Under these conditions, the capacity of the stairs saturates quickly, and so all floors have a slow evacuation. Figure \ref{figure3} shows a snapshot from one simulation of this strategy. Here, the profile of the queues at each level can be observed. The differences in the length of queues are due to differences in the temporal evolution of density in front of each door. 

In this evacuation scheme, the first level that can be emptied is the $1^{st}$ floor ($105 \pm 6$~s) and the last one is the $6^{th}$ floor ($259 \pm 3$ s).

The Total Evacuation Time (\textit{TET}) of the building for this configuration is $316 \pm 8$~s, and the mean Floor Evacuation Time (${FET}$) is $195 \pm 55$~s. 

For reference, the independent evacuation of a single floor toward the stairs was also simulated. It was found that the evacuation time of only one level toward the empty stair is $65 \pm 4$ s.

\subsection{Bottom-Up strategy}

\begin{figure*}%[th]
\psfrag{FET}[c][c]{${FET}$}
\psfrag{TET}[c][c]{\textit{TET}}
\begin{center}
\subfigure[]{\includegraphics[width=0.40\textwidth]{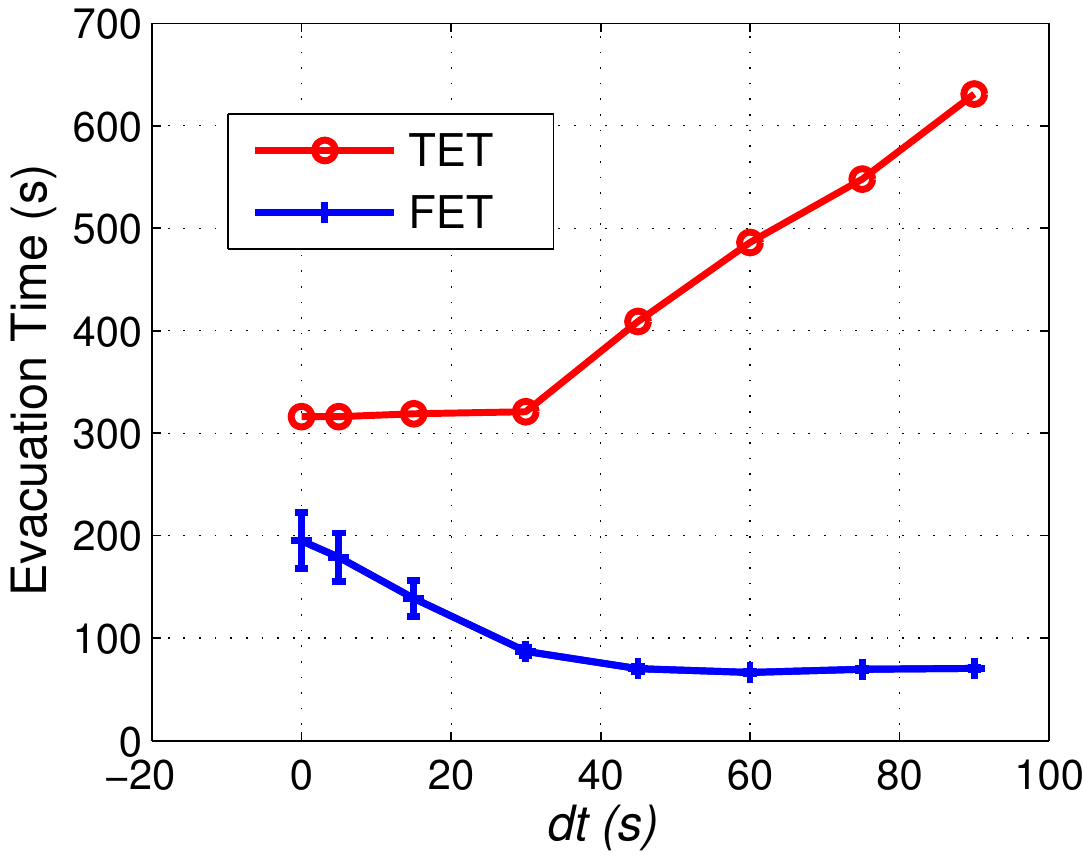}}
\hfil
\subfigure[]{\includegraphics[width=0.43\textwidth]{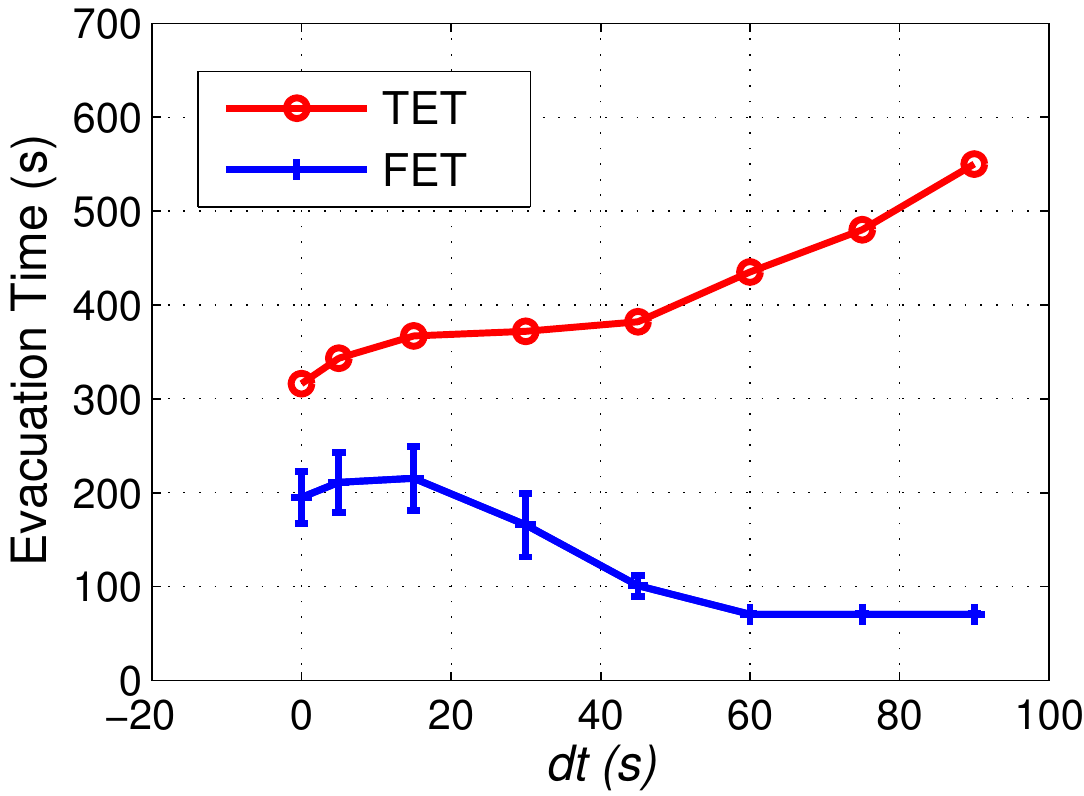}}
\end{center}
\caption{\textit{TET} and ${FET}$, obtained from simulations for different phase shifts ($dt$) following sequential evacuation: (a) Bottom-Up strategy, (b) Top-Down strategy. The symbols and error bars indicate one standard deviation.}\label{figure4}
\end{figure*}

Figure \ref{figure4}(a) shows the evacuation times for different time delays $dt$ following the Bottom-Up strategy. 

It can be seen that the Total Evacuation Time (\textit{TET}) remains constant for time delays ($dt$) up to 30 seconds. 
Therefore, \textit{TET} is the same as the simultaneous evacuation strategy ($dt = 0$ s) in this range. It is worth noting that 30 seconds is approximately one half of the time needed to evacuate a floor if the staircase were empty. 

Furthermore, the mean Floor Evacuation Time (${FET}$) declines as $dt$ increases, reaching the asymptotic value for $65$ seconds, which is the evacuation time of a single floor considering the empty stairway. 
As expected, if the levels are evacuated one at a time, with a time delay greater than the duration of the evacuation time of one floor, the system is at the limit of decoupled or independent levels. 
In these cases, \textit{TET} increases linearly with $dt$.

Since \textit{TET} is the same for $dt <  30$ s and ${FET}$ is significantly improved (it is reduced by half) for $dt = 30$ s, this phase shift can be taken as the best value, for this strategy, to evacuate this particular building. 

This result is surprising because the \textit{TET} of the building is not affected by systematic delays ($dt$)  at the start of the evacuation of each floor if $dt \leq 30$~s, which reaches up to 180 seconds for the floor that further delays the start of the evacuation. 

\begin{figure*}%[th]
\begin{center}
\includegraphics[width=0.95\textwidth]{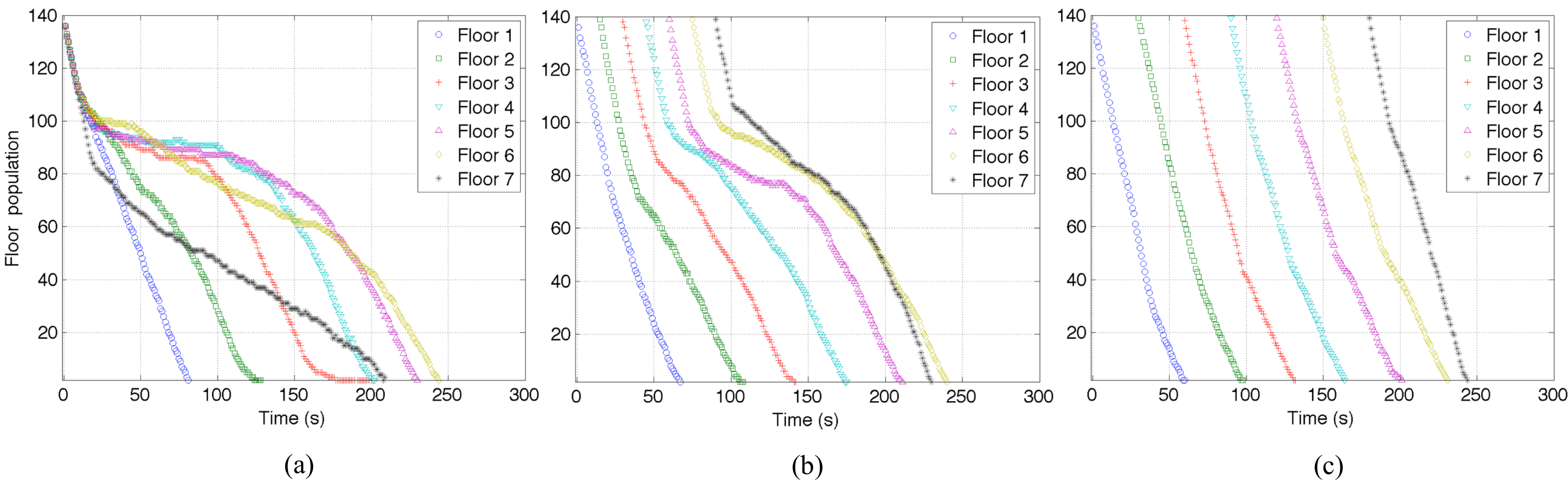}
\end{center}
\caption{Time evolution of the number of pedestrians in each floor up to 3 m before the exit to the staircase. (a) for the simultaneous evacuation ($dt=0$); (b) for delay of $dt=15$~s and (c) for $dt=30$~s.} 
\label{figurePop}
\end{figure*}

More details can be obtained by looking at the discharge curves corresponding to one realization of the building egress simulation. The evacuation of the first 140 pedestrians (93\%) of each floor is analyzed by plotting the occupation as a function of time in Fig. \ref{figurePop} for three time delays between the relevant range $dt~\epsilon[0, 30]$. For $dt=0$ [Fig. \ref{figurePop}(a)] there is an initial transient of about 10 seconds in which every floor can be evacuated toward a free part of the staircase before reaching the congestion due to the evacuation of lower levels. After that, it can be seen that the egress time of different floors has important variations,  the lower floors ($1^{st}$ and $2^{nd}$) being the ones that evacuate  quicker and intermediate floors  such as $5^{th}$ and $6^{th}$ the ones that take longer to evacuate. After an intermediate situation for $dt=15$~s [Fig. \ref{figurePop}(b)] we can observe the population profiles for the optimum phase shift of $dt=30$ 
in Fig. \ref{figurePop}(c). There, it can be seen that the first 140 occupants of different floors evacuate uniformly and very little perturbation from one to another is observed. 

In the curves shown in Fig. \ref{figurePop}, the derivative of the population curve is the flow rate, meaning that low slopes (almost horizontal parts of the curve such as the one observed in Fig. \ref{figurePop}(a) for the $5^{th}$ floor  between 40 and 100 s) can be identified with lower velocities and higher waiting time for the evacuating people. Because of the fundamental diagram, we know that lower velocities indicate higher densities. In consequence, we can  say that the greater the slope of the population curves, the greater the comfort of the evacuation (more velocity, less waiting time, less density). Therefore, it is clear that the situation displayed in Fig. \ref{figurePop}(c) is much more comfortable than the one in Fig. \ref{figurePop}(a).

In short,  for the Bottom-Up strategy, the time delay $dt=30$~s minimizes the perturbation  among  evacuating pedestrians from successive levels; it reduces  ${FET}$ to one half of the simultaneous strategy ($dt=0$~s); it maintains the total evacuation time (\textit{TET}) at the minimum and, overall, it exploits the maximum capacity of the staircase maintaining  each pedestrian's evacuation time at a minimum. This result is highly beneficial for the general system and for each floor, because it can avoid situations generating impatience due to waiting for gaining access to the staircase.

\subsection{Top-Down strategy}

Figure \ref{figure4}(b) shows the variation of \textit{TET} and ${FET}$, as a function of the time delay $dt$, for the Top-Down strategy. 

It must be noted that \textit{TET} increases monotonously for all $dt$, which is sufficient to rule out this evacuation scheme.

In addition, for $dt < 15$ s, ${FET}$ also increased, peaking at $dt = 15$ s. It can be said that for the system studied, the Top-Down strategy with a time delay of $dt = 15$ s leads to the worst case scenario. 

For $15$ s $< dt < 45$ s, there is a change of regime in which ${FET}$ decreases and \textit{TET} stabilizes. 

For values of $dt > 45$ s, ${FET}$ reaches the limit of independent evacuation of a single floor (see section \ref{sec:result}\ref{subsec:SimultaneousEvac}). And the TET of the building increases linearly due to the increasing delays between the start of the evacuation of the different floors. 

In summary, the Top-Down Strategy does not present any improvement with respect to the standard strategy of simultaneous evacuation of all floors ($dt = 0$).

%%%%%%%%%%%%%%%%%%%%%%%%%%%%%%%%%%%%%%%%%%%%%%%%%%%%%%%%%%%%%%%%%
\section{Conclusions}\label{sec:conclusions}

In this paper, we studied the evacuation of several pedestrian reservoirs (``rooms'') toward the same means of egress (``hallway''). In particular, we focused on an example, namely, a multistory building in which different floors are evacuated toward the staircase. We studied various strategies using computer simulations of people's movement. 

A new methodology, consisting in the sequential evacuation of the different floors (after a time delay $dt$) is proposed and compared to the commonly used strategy in which all the floors begin to evacuate simultaneously. 

For the system under consideration, the present study shows that if a strategy of sequential evacuation of levels begins with the evacuation of the $1^{st}$ floor and, after a delay of 30 seconds (in this particular case, $30$ s is approximately one half of the time needed to evacuate only one floor if the staircase were empty), it follows with the evacuation of the $2^{nd}$ floor and so on (Bottom-Up strategy), the quality of the overall evacuation process improves. From the standpoint of the evacuation of the building, \textit{TET} is the same as that for the reference state. However, if ${FET}$ is considered, there is a significant improvement since it falls to about half. This will make each person more comfortable during an evacuation, reducing the waiting time and thus, the probability of causing anxiety that may bring undesirable consequences. 

So, one important general conclusion is that a sequential Bottom-Up strategy with a certain phase shift can improve the quality of the evacuation of a building of medium height. 

On the other hand, the simulations show that the sequential Top-Down strategy is unwise for any time delay ($dt$). In particular, for the system studied, the value $dt = 15$ s leads to a very poor evacuation since the \textit{TET} is greater than that of the reference, and it maximizes ${FET}$ (which is also higher than the reference value at $dt = 0$). In consequence, the present study reveals that this would be a bad strategy that should be avoided.

The perspectives for future work are to generalize this study to buildings with an arbitrary number of floors (tall buildings), seeking new strategies.
We also intend to analyze strategies where some intermediate floor must be evacuated first (e.g., in case of a fire) and then the rest of the floors. 

The results of the present research could form the basis for developing new and innovative alarm systems and evacuation strategies aimed at enhancing the comfort and security conditions for people who must evacuate from pedestrian facilities, such us multistory buildings, schools, universities, and other systems in which several ``rooms'' share a common means of escape.

\begin{acknowledgements}
This work was financially supported by Grant PICT2011 - 1238 (ANPCyT, Argentina).
\end{acknowledgements}

\end{document}